\documentclass[conference]{IEEEtran}
\usepackage{color}

\ifCLASSINFOpdf
  \usepackage[pdftex]{graphicx}
\else
  \usepackage[dvips]{graphicx}
\fi
\begin{document}
%
\title{Mapping of the Quantum Efficiency of a Superconducting Single Electron
Detector}



%
\author{\IEEEauthorblockN{Adrian Lupa\c{s}cu\IEEEauthorrefmark{1},
Andreas Emmert\IEEEauthorrefmark{2},
Michel Brune\IEEEauthorrefmark{2},
Gilles Nogues\IEEEauthorrefmark{2},\\
Michael Rosticher\IEEEauthorrefmark{3},
Jean-Paul Maneval\IEEEauthorrefmark{3},
Francois-Rene Ladan\IEEEauthorrefmark{3}\\ and
Jean-Claude Villegier\IEEEauthorrefmark{4}}
\IEEEauthorblockA{\IEEEauthorrefmark{1}Laboratoire Kastler Brossel - ENS, CNRS, UPMC, 24 rue Lhomond, F-75231 Paris, France\\\emph{Present address}: IQC, University of Waterloo, 200 University Av W, Waterloo ON N2L 3G1, Canada\\ Email: alupascu@iqc.ca}
\IEEEauthorblockA{\IEEEauthorrefmark{2}Laboratoire Kastler Brossel - ENS, CNRS, UPMC, 24 rue Lhomond, F-75231 Paris, France}
\IEEEauthorblockA{\IEEEauthorrefmark{3}Laboratoire Pierre Aigrain - ENS, CNRS, UPMC, 24 rue Lhomond, F-75231 Paris, France}
\IEEEauthorblockA{\IEEEauthorrefmark{4}CEA,INAC-SPSMS, 17 rue des Martyrs, 38054 Grenoble CEDEX-9, France}}


\maketitle

\begin{abstract}
Superconducting NbN wires have recently received attention as detectors for visible and infrared photons~\cite{goltsman_2001_1}. We present experiments in which we use a NbN wire for high-efficiency ($\simeq 40\%$) detection of single electrons with keV energy. We use the beam of a scanning electron microscope as a focussed, stable, and calibrated electron source. Scanning the beam over the surface of the wire provides a map of the detection efficiency. This map shows features as small as 150~nm, revealing wire inhomogeneities. The intrinsic resolution of this mapping method, superior to optical methods, provides the basis of a characterization tool relevant for photon detectors.
\end{abstract}


%
\IEEEpeerreviewmaketitle

\section{Introduction}
%
%

\let\thefootnote\relax\footnotetext{\copyright 2009 EEE.  Reprinted, with permission, from Proceedings of the 2009 IEEE Toronto International Conference, Science and Technology for Humanity (TIC-STH), page 1011-1014, 2009.

This material is posted here with permission of the IEEE.  Internal or
personal use of this material is permitted.  However, permission to
reprint/republish this material for advertising or promotional purposes
or
for creating new collective works for resale or redistribution must be
obtained from the IEEE by writing to pubs-permissions@ieee.org.

By choosing to view this document, you agree to all provisions of the
copyright laws protecting it.}

Superconducting single-photon detectors (SSPD)~\cite{semenov_2001_1,goltsman_2001_1} for the visible and infrared ranges have recently received a lot of attention due to their high efficiency, low reset time, and photon number discrimination capability~\cite{goltsman_2007_miv,divochiy_2008_snp}. These detectors use a superconducting nanowire, possibly shaped into a meander for larger collection area, biased with a current $I$ slightly lower than the critical current $I_{c}$. An incident photon absorbed by the wire creates a hot region which triggers a transition that can be detected as a voltage pulse.

We describe experiments in which we apply this principle to the detection of single electrons with keV energies. We note that hot-spot creation by single charged particles (MeV-energy $\alpha$ particles) was observed indirectly as a reduction of the critical current of a superconducting slab in~\cite{spiel_1965_alpha-det}. In our experiments we focus a calibrated electron beam on a superconducting nanowire and we count wire transition events. We detect single electrons with an efficiency as high as $\mbox{40}\,\%$. Here we present experiments in which we take advantage of the tightly focussed electron beam to test the local single-particle response of the detector. We note that our work differs from previous low-temperature scanning-electron microscopy (LTSEM) studies of superconducting films~\cite{gross_1994_lts,doenitz_2007_1} that focussed on response in the stationary regime. Using single-electron detection we observe wire inhomogeneities with a spatial resolution considerably improved with respect to both previous LTSEM work and also with optical scanning methods~\cite{hadfield_2007_res-sspd}.

\section{Experiments}

\subsection{Device fabrication}

Our detector devices are realized starting with a 10~nm thick NbN film sputtered epitaxially on a MgO~(100) substrate~\cite{villegier_2001_NbN-sapphire}. The critical temperature of the film is $T_{c}=14\,\mbox{K}$. A PMMA resist layer is applied and patterned using electron-beam lithography. Reactive ion etching with SF$_{\mbox{6}}$ is used to define the wire geometry. Finally a Au layer with a thickness of $\mbox{5}\;\mbox{nm}$ is evaporated. We present measurements on two samples, \emph{A} and \emph{B}, with a width of the wire of $0.5\,\mu\mbox{m}$ and  $1\,\mu\mbox{m}$ respectively, and a length of $100\,\mu\mbox{m}$.

\subsection{Experimental setup}

\begin{figure}[!t]
\centering
\includegraphics[width=3.2in]{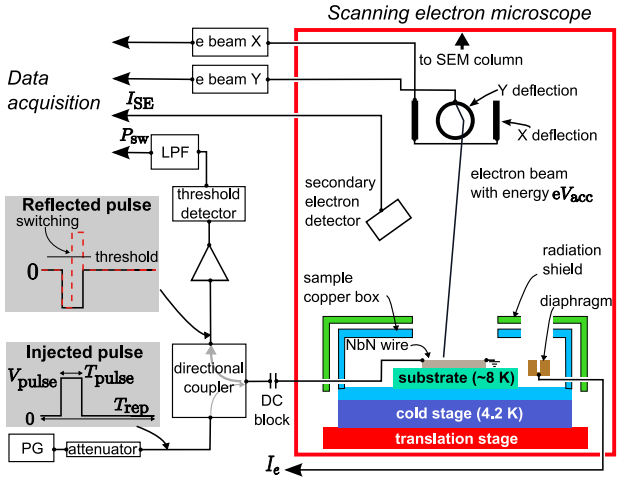}
\caption{ Experimental setup for the characterization of electron detection.}
\label{fig1}
\end{figure}

The characterization of electron detection efficiency requires the use of a calibrated and stable electron source. In addition, a focussed source provides the possibility to test this detection efficiency locally. The ideal electron source for our experiment is therefore a scanning electron microscope (SEM). We use a JEOL JSM-840 SEM. The detector sample is cooled using an Oxford Instruments CF-313 4He circulation cryostat, whose cold stage is designed to be mounted on the translation stage of the SEM (see Fig.~\ref{fig1}). The detector wire is connected using wire bonding to contact pads on a printed circuit board, which further lead to coaxial connectors. The use of a radiation shield and good thermal contact between the substrate and the copper box are essential for reducing the sample temperature.

\begin{figure}[!t]
\centering
\includegraphics[width=3.2in]{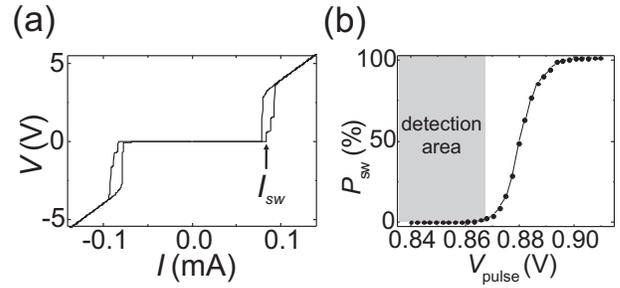}
\caption{\label{fig2} Characterization measurements for sample \emph{A} at $T=9\,\mbox{K}$. (a) The V-I characteristic of sample \emph{A}. The switching current $I_{sw}$ is indicated. (b) Switching probability $P_{\mbox{\small{sw}}}$ as a function of the voltage amplitude of $V_{\mbox{\small{pulse}}}$. The gray area indicates the range of $V_{\mbox{\small{pulse}}}$ usable for electron detection.}
\end{figure}

A first characterization of the sample is done by measuring the DC $V-I$ characteristics, using a current source and a voltmeter (this part of the setup is not shown in Fig.~\ref{fig1}). The result is shown in Fig.~\ref{fig2}(a) for sample \emph{A}. The switching current is $I_{\mbox{\small{sw}}}=86\,\mu\mbox{A}$ and the normal state resistance is $R_{\mbox{\small{n}}}=40.4\,\mbox{k}\Omega$. The voltage jumps observed for currents larger than $I_{\mbox{\small{sw}}}$ reveal the presence of phase-slip centers~\cite{skocpol_1974_psc}.

For electron detection, we use the setup shown in Fig.~\ref{fig1}. We bias the sample with electrical pulses of duration $T_{\mbox{\small{pulse}}}$, voltage amplitude $V_{\mbox{\small{pulse}}}$, and repetition period $T_{\mbox{\small{rep}}}$, generated by a pulse generator (PG). This biasing scheme allows us to decouple the sample using a DC block, and therefore reduces the low-frequency noise present in the system. The pulses are injected to the sample through a directional coupler, which allows us to recover the signal reflected from the sample at a third port. If the wire remains in the superconducting state through the duration of the pulse, its impedance is very low and the reflected signal has the same amplitude but opposite sign as compared to the incident pulse (continuous curve in Fig.~\ref{fig1}). A current-induced transition changes the impedance of the wire to a value significantly higher than $50\,\Omega$, which results in a change of sign in the reflected signal (dashed curve in Fig.~\ref{fig1}). The transition can be detected using a threshold detector (LeCroy 821), which provides an output digital signal for each switching event. Analog averaging of this output signal is done using a low-pass filter (LPF) with a time constant of $\approx 10^{3}\, T_{\mbox{\small{rep}}}$. This yields an accurate measurement of the switching probability $P_{\mbox{\small{sw}}}$. In Fig~\ref{fig2}(b) we plot $P_{\mbox{\small{sw}}}$ as a function of $V_{\mbox{\small{pulse}}}$, without electrons, for $T_{\mbox{\small{pulse}}}=\mbox{100 ns}$ and a repetition time $T_{\mbox{\small{rep}}}=\mbox{3 }\mu\mbox{s}$ chosen long enough to ensure full thermal recovery of the superconducting film. We find that $P_{\mbox{\small{sw}}}=50\,\%$ for a pulse amplitude $V_{\mbox{\small{pulse}}}^{50\,\%}=0.88\,\mbox{V}$, corresponding to a current amplitude at the sample of $122\,\mu\mbox{A}$, as estimated by taking into account the attenuation in the coaxial cables leading to the sample. The discrepancy between this value and the DC switching current is probably due to the additional noise present while acquiring the DC characteristic and to thermal fluctuation effects~\cite{engel_2006_fes}.

\subsection{Characterization of the electron detection efficiency}

For the characterization of electron detection, we set $V_{\mbox{\small{pulse}}}$ to a value such that $P_{\mbox{\small{sw}}}$ is very low, typically less than $1\,\%$. This ensures that the dark count rate of the detector is negligible. The electron beam is then switched on and scanned over an area of $20\,\mu \mbox{m}\, \mbox{x}\, 120\,\mu \mbox{m}$ that contains the wire. During this scan the $X$ and $Y$ position of the electron beam, as well as the SEM secondary electron detector signal $I_{\mbox{\small{SE}}}$ and the switching probability $P_{\mbox{\small{sw}}}$, are recorded. The electrons are accelerated at a voltage $V_{\mbox{\small{acc}}}$ and the beam current is set to a value in the pA range. A reliable measurement of the beam current is realized using a diaphragm mounted in the copper box close to the sample ($I_{\mbox{\small{e}}}$). The switching probability of the wire is given by $P_{\mbox{\small{sw}}}=1-\exp{(-\Gamma_{\mbox{\small{sw}}}T_{\mbox{\small{pulse}}})}$, with the switching rate $\Gamma_{\mbox{\small{sw}}}=\epsilon \phi_{\mbox{\small{e-beam}}}$, where $\phi_{\mbox{\small{e-beam}}}$ is the electron flux and $\epsilon$ is the quantum detection efficiency. The electron flux is calculated as $\phi_{\mbox{\small{e-beam}}}=I_{\mbox{\small{e}}}/e$. The measurement of $P_{\mbox{\small{sw}}}$ and $I_{\mbox{\small{e}}}$ is thus used to calculate the quantum efficiency $\epsilon$.

\begin{figure}[!t]
\centering
\includegraphics[width=3.2in]{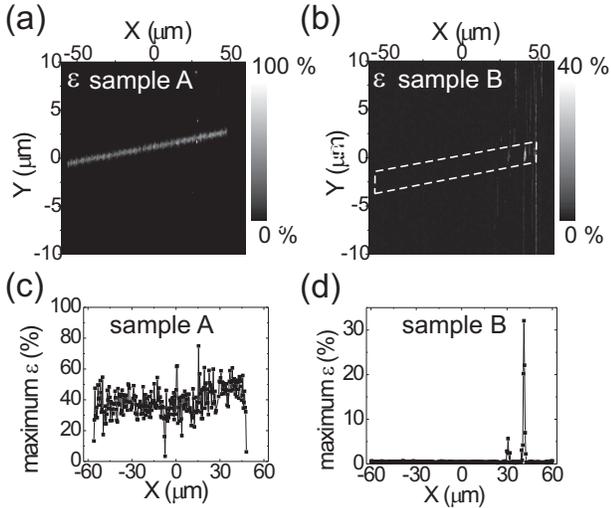}
\caption{\label{fig3} Electron detection efficiency maps (a,b) and maximum detection efficiency along the wire (c,d) for samples \emph{A} ($V_{\mbox{\small{acc}}}=5\,\mbox{kV}$ and $I_{\mbox{\small{e}}}=2\,\mbox{pA}$) and \emph{B} ($V_{\mbox{\small{acc}}}=5\,\mbox{kV}$ and $I_{\mbox{\small{e}}}=1.2\,\mbox{pA}$) respectively. The dashed contour in (b) is an eye-guide to the borders of the $100\,\mu\mbox{m}\,\times\,1\,\mu\mbox{m}$ wire.}
\end{figure}

In Fig.~\ref{fig3}(a) we show a gray-scale plot of the quantum efficiency $\epsilon$ for sample \emph{A}, recorded during a single scan with an electron beam of energy $V_{\mbox{\small{acc}}}=5\,\mbox{kV}$ and current $I_{\mbox{\small{e}} }=2\,\mbox{pA}$ (corresponding to $1.25$ electrons reaching the wire during $T_{\mbox{\small{pulse}}}$). We find a significant value of the detection efficiency $\epsilon$ only over the wire surface, as indicated by the coincidence with the regular SEM image (signal $I_{\mbox{\small{SE}}}$, not shown). For the chosen value of the amplitude of the pulses $V_{\mbox{\small{pulse}}}=\mbox{97}\%\, \times V_{\mbox{\small{pulse}}}^{50\,\%}$, and for $2\;\mbox{pA}\leq I_{\mbox{\small{e}}}\leq8.5\;\mbox{pA}$, the number of detection events is found to depend linearly upon the electron flux $I_{\mbox{\small{e}}}$. This is a clear indication of single-particle detection, as also supported by the estimation of the thermal recovery time of the NbN film. This time is found to be much shorter than the time interval between two electrons (ranging between 80 and 19~ns for the current range above). Indeed, from the phonon escape time~\cite{semenov_2001_1} and the heat conduction of the MgO crystal itself~\cite{grimwall_1986_tpm} one concludes that the NbN film has resumed to the quiescent temperature in less than 2~ns after the arrival of one single electron.

In order to quantify the detection efficiency along the wire, we extract the maximum efficiency for each scanning line (such a line corresponds to a fixed $X$ value in Fig.~\ref{fig3}(a)). The result, shown in Fig.~\ref{fig3}(c), indicates a weak variation, except for a few pronounced peaks and dips; the average efficiency is $40\;\%$. In Fig.~\ref{fig3}(b) and (d) we show similar measurements of the detection map and maximum efficiency along wire respectively for sample \emph{B}. In contrast to the case of sample \emph{A}, only a few regions have a significant quantum efficiency. We attribute the observed efficiency variations to inhomogeneities of the wires. For sample \emph{B}, we find that the DC switching current is $102\,\mu\mbox{A}$, only slightly higher than for sample \emph{A}, for a nominal width twice as large. This indicates the presence of a weak spot, which limits the current that can be applied before switching. This is likely the cause of the low detection efficiency for sample \emph{B}.

\begin{figure}[!t]
\centering
\includegraphics[width=3.1in]{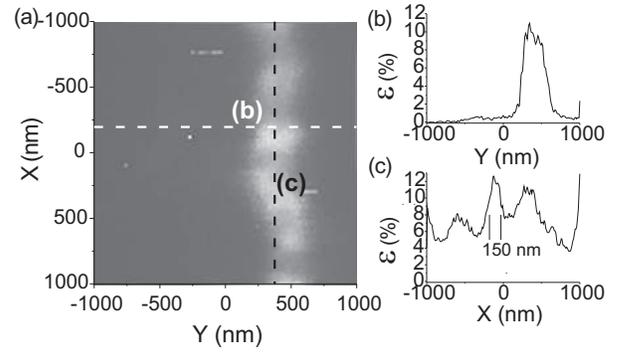}
\caption{\label{fig4} (a) Efficiency map for sample \emph{A} for $V_{\mbox{\small{acc}}}=20\,\mbox{kV}$, $I_{\mbox{\small{e}}}=13.3\,\mbox{pA}$, and $V_{\mbox{\small{pulse}}}=\mbox{97}\%\, \times V_{\mbox{\small{pulse}}}^{50\,\%}$. (b,c) Efficiency along the dashed lines indicated in (a).}
\end{figure}

A high-resolution map of the detection efficiency for a $2\,\mu\mbox{m}\,\times\,2\,\mu\mbox{m}$ area is shown in Fig.~\ref{fig4}(a) for sample A. This measurement is done at $V_{\mbox{\small{acc}}}=20\,\mbox{kV}$, where the spatial resolution is found to be higher than at 5~kV, despite the detection efficiency being reduced. The cross-section perpendicular to the wire of Fig.~\ref{fig4}(b) shows a sensitive region with an extent of 350~nm (FWHM). This is smaller than the lithographically defined width of the wire of 500~nm, revealing an inhomogeneous detection area. The cross-section along the wire (Fig.~\ref{fig4}(c)) displays efficiency variations over length scales as short as 150~nm. We note that the measured map is a convolution of the local wire properties with the finite spatial resolution of our method. These results give an upper limit to the latter of 150 nm.

\section{Conclusions and perspectives}
Our results show that an SEM can be used as a local probe for wire inhomogeneities that determine single-particle detection efficiency. We note that the achieved resolution is superior to the resolution achieved in recent experiments, when the single \emph{photon} response was studied~\cite{hadfield_2007_res-sspd}. Therefore we expect this type of probing to be relevant for optimizing the properties of SSPDs.

High-efficiency superconducting single-electron detectors are potentially interesting for applications that impose working in a low-temperature and low-dissipation environment. Our goal is the development of such a detector for counting single Rydberg atoms~\cite{gallagher_1994_rydbergs}. We are interested in the measurement of atomic circular Rydberg states~\cite{gallagher_1994_rydbergs}, for which electron detection with the constraints mentioned above is necessary. To distinguish two circular Rydberg states, an electric field is applied such that ionization occurs predominantly for one of the two states~\cite{gallagher_1994_rydbergs}; then the electron resulting from ionization has to be detected. An electron detector for measurement of Rydberg state which is compatible with preservation of atomic state coherence has to function at low temperatures and have low dissipation.  For a new class of experiments, in which Rydberg atoms are produced and manipulated on atom chips~\cite{hyafil_2004_2,nirrengarten_2006_1}, a compact detector, ideally integrated with the atom chip, would be most suitable.


\section*{Acknowledgment}
We thank J.-M. Raimond, S. Haroche, D. Est$\grave{\mbox{e}}$ve, D. Vion, and P. Bertet for useful discussions. D.E. and D.V. also lent us the cryostat used in these experiments. This work was supported by the Cnano IdF, the SINPHONIA-NMP4 project, the SCALA project of EU, and the SBPC ANR program. AL acknowledges support from the EU through the Marie Curie program.



\bibliographystyle{IEEEtran}

%



\end{document}